\documentclass[11pt]{article}
\usepackage{amsmath,amssymb,amsthm}
\usepackage[margin=3cm]{geometry}

\numberwithin{equation}{section}

\begin{document}

\title{A Vaidya-type generalisation of Kerr space-time}

\author{C. G. B\"ohmer\footnote{Email: c.boehmer@ucl.ac.uk}, \\
Department of Mathematics, University College London,\\
Gower Street, London WC1E 6BT, United Kingdom,\\
and\\P. A. Hogan\footnote{Email: peter.hogan@ucd.ie}, \\
School of Physics, University College Dublin,\\
Belfield, Dublin 4, Ireland}

\date{}

\maketitle

\begin{abstract}
A new Vaidya-type generalisation of Kerr space-time is constructed by requiring the Kerr mass and angular momentum per unit mass to depend upon a variable which has a simple geometrical origin. The matter distribution introduced in this way radiates mass and angular momentum at future null infinity. The Vaidya generalisation of the Schwarzschild space-time is a special case of the newly found solution. 
\end{abstract}

\section{Introduction}

Shortly after the formulation of the Einstein field equations, Schwarzschild was able to derive an exact static and spherically symmetric solution of the gravitational field equations. To date, this is still viewed as probably the most important known solution. The Schwarzschild line element in outgoing Eddington--Finkelstein coordinates reads
\begin{align}
  \label{1}
  ds^2=-\left (1-\frac{2\,m}{r}\right )du^2-2\,du\,dr+r^2(d\theta^2+\sin^2\theta\,d\phi^2)\ ,
\end{align}
where $m$ is the mass parameter of the solution which is assumed to be constant.

About 25 years later Vaidya was able to generalise the Schwarzschild space-time such that the space-time would radiate mass to future null infinity. This solution is sometimes referred to as the radiating Schwarzschild metric. The Vaidya line element (originally given in \cite{V}) no longer relies on the assumption of a constant mass parameter and considers $m=m(u)$. If $dm/du\neq 0$ then we no longer have a vacuum space-time since the Ricci tensor, in coordinates $x^i=(u,r,\theta, \phi)$, has components $R_{ij}$ given by
\begin{align}
  \label{2}
  R_{ij}=\frac{2\,\dot m}{r^2}\,k_i\,k_j\ ,
\end{align}
with $\dot m=dm/du$ and $k_i\,dx^i=du$ defining a future pointing radial null vector field. 

The approach to looking for a Vaidya-type generalisation of the Kerr space-time has focused on constructing a line element which generalises the Kerr line element and which satisfies Einstein's non-vacuum field equations with an energy-momentum-stress tensor which is proportional to the square of a null vector field, as in the case of (\ref{2}). When this tensor vanishes, the Kerr solution is recovered. The resulting current situation is neatly summarised in \cite{Exact} as follows: ``In the axisymmetric case, the complete solution was first found by Herlt \cite{H}, using a formalism developed by Vaidya \cite{V1,V2}''. The Herlt solution includes the so-called ``radiating Kerr metric'' constructed by Vaidya and Patel \cite{V3}. It is also importantly emphasised in \cite{Exact} that none of the non-vacuum solutions of this type can be interpreted as having a pure radiation Maxwell field as source. The Vaidya line element continues to stimulate research as evidenced by a selection of recent papers \cite{b1,b2,b3,b4} on the topic. 

In this paper we are taking a different approach. Instead of looking for a Vaidya-type generalisation of Kerr space-time as described above, we begin by asking the following question: If the Kerr parameters of mass $m$ and angular momentum per unit mass $a$ are no longer constants, then can we find a natural independent variable that they can depend upon, having the property that if the variable angular momentum per unit mass vanishes then the line element reduces to the Vaidya generalisation of the Schwarzschild space-time?

The answer to this question is affirmative and relies on a simple observation. Namely, one can work with a new variable $u=t-r$ where $r$ is now the radial coordinate related to the Boyer--Lindquist coordinates. We will see that in our approach this choice appears rather naturally.

\section{Spherically symmetry -- radiating Schwarzschild solution}
\label{sec:schw}

In the following we will briefly recall the main results of the Vaidya generalisation of the Schwarzschild space-time which are required when considering the Kerr space-time. Let us begin with the Kerr-Schild form of the Schwarzschild line element in rectangular Cartesian coordinates and time $X^i=(x, y, z, t)$ given by
\begin{align}
  \label{3}
  ds^2=-dt^2+dx^2+dy^2+dz^2+\frac{2\,m}{r}(k_i\,dX^i)^2=g_{ij}\,dX^i\,dX^j\ ,
\end{align}
with $m$ being the constant mass parameter. Note that the line element (\ref{1}) is in Kerr-Schild form in coordinates $x^i=(u,r,\theta, \phi)$. The null vector $k_i$ is given by
\begin{align}
  \label{4}
  k_i\,dX^i=-dt+\frac{x}{r}dx+\frac{y}{r}dy+\frac{z}{r}dz\ ,
\end{align}
where the radius $r$ is given by its standard Euclidean distance from the origin
\begin{align}
  \label{5}
  r^2=x^2+y^2+z^2\ .
\end{align}
We see from (\ref{3}) that
\begin{align}
  \label{6}
  g_{ij}=\eta_{ij}+\frac{2\,m}{r}k_i\,k_j\ , 
\end{align}
with $\eta_{ij}={\rm diag}(-1, +1, +1, +1)$ and (\ref{4}) defines a null vector field with respect to the metric with components $g_{ij}$ and with respect to the Minkowskian metric with components $\eta_{ij}$. Moreover, $k^i=g^{ij}\,k_j=\eta^{ij}\,k_j$ with $g^{ij}, \eta^{ij}$ defined by $g^{ij}\,g_{jk}=\delta^i_k=\eta^{ij}\,\eta_{jk}$. 

Now we make the crucial observation that
\begin{align}
  \label{7}
  g_{ij}\,k^i\,X^j=\eta_{ij}\,k^i\,X^j=-t+r\ .
\end{align}
It is this relation which will allow us to construct a new Vaidya-type generalisation of the Kerr metric by considering this equation in the axisymmetric case.

Let $u=t-r$ so that $u$ increases when $t$ increases (outgoing Eddington--Finkelstein) then we can derive the following relation
\begin{align}
  \label{8}
  u_{,i}=-k_i\ ,
\end{align}
where the comma denotes partial differentiation with respect to $X^i$. Hence $u={\rm constant}$ are null hypersurfaces. With this function $u$ we generalise metric (\ref{3}) to 
\begin{align}
  \label{9}
  ds^2=-dt^2+dx^2+dy^2+dz^2+\frac{2\,m(u)}{r}(k_i\,dX^i)^2=g_{ij}\,dX^i\,dX^j\ ,
\end{align}
with $\dot m=dm/du\neq 0$. Introducing standard spherical polar coordinates $x=r\,\sin\theta\,\cos\phi, y=r\,\sin\theta\,\sin\phi, z=r\,\cos\theta$ and writing (\ref{9}) in terms of coordinates $(u,r,\theta, \phi)$ transforms (\ref{9}) into (\ref{1}) but with non-constant mass.

\section{Axisymmetric case -- radiating Kerr}

\subsection{Basic quantities}

Let us begin, as above, with the Kerr solution of Einstein's vacuum field equations in Kerr--Schild form so that the metric tensor components, in coordinates $X^i=(t,x, y, z)$, have the 
form
\begin{align}
  \label{10}
  g_{ij}=\eta_{ij}+2\,H\,k_i\,k_j\ ,
\end{align}
where the function $H$ is given by 
\begin{align}
  \label{11}
  H=\frac{m\,r^3}{r^4+a^2z^2}\ .
\end{align}
The mass of the source of denoted by $m$ while the source angular momentum per unit mass is denoted by $a$. Both quantities are assumed to be constants. The components $k_i$ are given via the 1--form
\begin{align}
  \label{12}
  k_i\,dX^i = -dt+
  \left(\frac{r\,x+a\,y}{r^2+a^2}\right ) dx +
  \left (\frac{r\,y-a\,x}{r^2+a^2}\right ) dy +
  \frac{z}{r}\,dz\ .
\end{align}
Here the radial coordinate $r$ is a function of $x, y, z$ given by the equation
\begin{align}
  \label{13}
  \frac{x^2+y^2}{r^2+a^2}+\frac{z^2}{r^2}=1\ .
\end{align}
The Boyer--Lindquist coordinates satisfy this relation. We note that $k_i=\eta_{ij}\,k^j$ is null with respect to the metric $\eta_{ij}$ (and thus null with respect to the metric $g_{ij}$) and $g^{ij}=\eta^{ij}-2\,H\,k^i\,k^j$ is the inverse of $g_{ij}$.

To find a natural independent variable upon which the mass $m$ and the angular momentum per unit mass $a$ should depend, to provide a Vaidya type generalisation of the Kerr space--time, we proceed as in the Schwarzschild case. Using (\ref{10}), (\ref{12}) and (\ref{13}) we find that
\begin{align}
  \label{14}
  g_{ij}\,k^i\,X^j = \eta_{ij}\,k^i\,X^j = 
  -t+\frac{(r\,x+a\,y)\,x+(r\,y-a\,x)\,y}{r^2+a^2} + 
  \frac{z^2}{r}=-t+r\ .
\end{align}
This is identical in form to (\ref{7}) but, of course, the coordinate $r$ in (\ref{7}) is given in terms of the rectangular Cartesian coordinates $x, y, z$ by (\ref{5}) whereas the coordinate $r$ in (\ref{14}) is given in terms of $x, y, z$ by (\ref{13}). More geometrically speaking, the surfaces of constant $r$ given by (\ref{5}) are spheres while the surfaces of constant $r$ given by (\ref{13}) are ellipsoids or more precisely oblate spheroids. 

Following the pattern in Section~\ref{sec:schw} we put $u=t-r$ in this case and consider a Vaidya generalisation of the Kerr space-time to be given by (\ref{10})--(\ref{13}) with $m=m(u)$ and $a=a(u)$. 

From (\ref{13}) with $a=a(u)$ and $u=t-r$ we find that the partial derivatives of $r$ with respect to $X^i$, indicated by a comma, are given by
\begin{align}
  \label{15}
  r_{,i}=\left(-\frac{a\,\dot a\,r\,(r^2-z^2)}{D},\frac{r^3\,x}{D}, \frac{r^3\,y}{D}, \frac{r\,z\,(a^2+r^2)}{D}\right )\ ,
\end{align}
where the denominator $D$ is
\begin{align}
  \label{16}
  D=r^4+a^2\,z^2-a\,\dot a\,r\,(r^2-z^2)\ .
\end{align}
Consequently, one can compute
\begin{align}
  \label{17}
  u_{,i}= \left(
  \frac{r^4+a^2\,z^2}{D},
  -\frac{r^3\,x}{D},
  -\frac{r^3\,y}{D}, 
  -\frac{r\,z\,(a^2+r^2)}{D}
  \right)\ ,
\end{align}
and one verifies that the following holds
\begin{align}
  \label{18}
  u_{,i}\,k^i=0\ .
\end{align}
Moreover, one can also show that 
\begin{align}
  \label{19}
  g^{ij}\,u_{,i}\,u_{,j} =\eta^{ij}\,u_{,i}\,u_{,j} =
  \frac{a^2\,(r^2-z^2)(r^4+a^2\,z^2)}{D^2}\ .
\end{align}

We thus see that $u={\rm constant}$ are \emph{not} null hypersurfaces in general, if $a\neq 0$. However, they are asymptotically null in the sense that, for large positive values 
of $r$, 
\begin{align}
  \label{20}
  g^{ij}\,u_{,i}\,u_{,j} = \frac{a^2}{r^2}\left (1-\frac{z^2}{r^2}\right ) +
  O\left (\frac{1}{r^3}\right ) = O\left (\frac{1}{r^2}\right )\ .
\end{align}
To see this one notes that $D^2 \sim r^8$ for large $r$ while the numerator of (\ref{19}) grows like $r^6$ for large radii. 

The future-pointing vector field given by the 1--form (\ref{13}), now with $a=a(u)$, is of course null and satisfies
\begin{align}
  \label{21}
  g^{ij}\,k_i\,k_j=\eta^{ij}\,k_i\,k_j=0\ .
\end{align}
It is also geodesic, this means it satisfies the geodesic equation
\begin{align}
  \label{22}
  k^i{}_{;j}\,k^j=k^i{}_{,j}\,k^j=0\ ,
\end{align}
where the semicolon denotes covariant differentiation with respect to the Riemannian connection calculated with the metric tensor $g_{ij}$. This null, geodesic vector field $k^i$ has expansion
\begin{align}
  \label{23}
  \frac{1}{2}k^i{}_{;i} = \frac{1}{2}k^i{}_{,i} = 
  \frac{2\,r^3-a\,\dot a\,r\,(r^2-z^2)}{D}\ ,\end{align}
and the squared modulus of its complex shear $\sigma$ is given by
\begin{align}\label{24} 
|\sigma |^2=\frac{1}{2}k_{(i;j)}\,k^{i;j}-\left (\frac{1}{2}\,k^i{}_{;i}\right )^2=\frac{1}{2}k_{(i,j)}\,k^{i,j}-\left (\frac{1}{2}\,k^i{}_{,i}\right )^2=\frac{a^2\dot a^2(r^2-z^2)^2}{4\,D^2}\ ,\end{align}
with the round brackets enclosing indices denoting symmetrisation. For large positive values of $r$ we find that
\begin{align}
  \label{25}
  u_{,i} = -k_i+O\left (\frac{1}{r}\right ),\quad \frac{1}{2}k^i{}_{;i} =
  \frac{1}{r}+\left (\frac{1}{r^2}\right ) \quad {\rm and}\quad |\sigma| = 
  O\left (\frac{1}{r^2}\right ),
\end{align} 
from which we can conclude that asymptotically the hypersurfaces $u={\rm constant}$ are future null cones, generated by expanding, shear-free null geodesics. 

\subsection{Ricci tensor, Ricci scalar and Einstein tensor}

Explicit computations involving axisymmetric space-times can be quite involved. In the following we note some useful expressions in the context of the metric 
\begin{align}
  \label{eqn:n1}
  g_{ij}=\eta_{ij}+2 \mathcal{H}_{ij}\ ,
\end{align}
where $\mathcal{H}_{ij}$ is given by $\mathcal{H}_{ij} = H k_i k_j$. We assume that $k^i$ is null and geodesic. Firstly, the Ricci tensor turns out to be linear in $\mathcal{H}_{ij}$. Let us define the object
\begin{align}
  \label{eqn:n2}
  \hat{\Gamma}^i_{jk} = 
  \mathcal{H}^i{}_{j,k} + \mathcal{H}^i{}_{k,j} - \eta^{is} \mathcal{H}_{jk,s}\ ,
\end{align}
then the Ricci tensor can be written as
\begin{align}
  \label{eqn:n3}
  R_{jk} = \hat{\Gamma}^s_{jk,s} =
 \mathcal{H}^s{}_{j,ks} + \mathcal{H}^s{}_{k,js} - \Box \mathcal{H}_{jk} \ .
\end{align}
Consequently, the Einstein tensor takes the following compact form 
\begin{align}
  G_{ij} = \Box \mathcal{H}_{ij} - 
  \mathcal{H}^s{}_{j,si} - \mathcal{H}^s{}_{i,sj} -
  \Bigl(\mathcal{H}_{ij} + \frac{1}{2} \eta_{ij}\Bigr) R \,,
\end{align}
where $R$ denotes the Ricci scalar. It should be noted that only the term $R \mathcal{H}_{ij}$ contains quadratic contributions in the function $H$.

When considering an asymptotic expansion for large radii, we note that the Ricci scalar decays like $1/r^6$ while the second partial derivatives of $\mathcal{H}_{ij}$ have leading terms that decay like $1/r^2$. In case of the Vaidya generalisation of the Schwarzschild metric one has $R=0$ which follows immediately from (\ref{2}). In the axisymmetric case $R \neq 0$, however, asymptotically the terms of the Ricci scalar do not contribute to the Einstein tensor. 

\subsection{Energy-momentum-stress tensor}

With the metric tensor given by (\ref{10})--(\ref{13}) with $m=m(u)$ and $a=a(u)$ we calculate the energy--momentum--stress tensor $T_{ij}$ of the matter distribution from Einstein's field equations
\begin{align}
  \label{26}
  -8\,\pi\,T_{ij} = G_{ij} \ .
\end{align}
We shall require $T_{ij}$ with sufficient accuracy to enable us to calculate the asymptotic flux of the 4--momentum
\begin{align}
  \label{27}
  P^i = \lim_{r\rightarrow+\infty}
  r^2\int_{u_0}^{u_1}du\int T^{ij}\,r_{,j}\,\sin\theta\,d\theta\,d\phi\ ,
\end{align}
and the asymptotic flux of angular momentum
\begin{align}
  \label{28}
  S^{ij} = \lim_{r\rightarrow+\infty}
  r^2\int_{u_0}^{u_1}du \int (T^{ki}\,X^j-T^{kj}\,X^i)\,r_{,k}\,\sin\theta\,d\theta\,d\phi\ ,
\end{align}
crossing $r={\rm constant}\rightarrow+\infty$ \emph{outwards} (in the direction of increasing $r$) between $u=u_0$ and some $u=u_1>u_0$ (say).

Here the polar angles $\theta, \phi$ arise from the parametrisation of  (\ref{13}) given by
\begin{align}
  \label{28'}
  x = \sqrt{r^2+a^2}\,\sin\theta\cos\phi\ , \quad
  y = \sqrt{r^2+a^2}\,\sin\theta\sin\phi\ , \quad
  z = r\,\cos\theta\ ,
\end{align}
with $0\leq\theta\leq\pi$ and $0\leq\phi\leq2\,\pi$. This parametrisation is the one used when working with Boyer--Lindquist coordinates. Detailed derivations of the 3--volume elements appearing here can be found, for example, in \cite{Synge}. Alternatively $P^i(u)$ satisfies the equation
\begin{align}
  \label{29}
  \frac{dP^i}{du} = \lim_{r\rightarrow+\infty}
  r^2\int T^{ij}\,r_{,j}\,\sin\theta\,d\theta\,d\phi\ ,
\end{align}
and the asymptotic flux of angular momentum $S^{ij}(u)$ satisfies 
\begin{align}
  \label{30}
  \frac{dS^{ij}}{du} = \lim_{r\rightarrow+\infty}
  r^2\int (T^{ki}\,X^j-T^{kj}\,X^i)\,r_{,k}\,\sin\theta\,d\theta\,d\phi\ .
\end{align}

Only part of the energy-momentum-stress tensor calculated from (\ref{26}) contributes to the quantities given by (\ref{29}) and (\ref{30}). This follows from the asymptotic behaviour for large radii. Denoting that part by $\hat T^{ij}$, an involved calculation reveals that it is given by
\begin{align}
  \label{31}
  8\,\pi\,\hat T^{ij} = -\frac{2\,\dot m}{r^2}\,\hat k^i\,\hat k^j -
  \frac{3}{r^3}\frac{d}{du}(m\,a)\,(\hat k^i\,\hat\lambda^j+\hat k^j\,\hat\lambda^i)\ .
\end{align}
The vectors $\hat k^i$ and $\hat\lambda^i$ are given by the following expressions
\begin{align}
  \label{32}
  \hat k^i=\left (1,\frac{x}{r}, \frac{y}{r}, \frac{z}{r}\right )\quad {\rm and}\quad
  \hat\lambda^i=\left (0,\frac{y}{r}, -\frac{x}{r}, 0\right )\ .
\end{align}
The spatial part of $\hat k^i$ is a unit normal vector, the vector $\hat\lambda^i$ is perpendicular to it.

For large positive values of $r$ we have (see also Section~\ref{sec:disc} below) $r_{,i}=\hat k_i-\hat v_i$ with $\hat v^i=\delta^i_0\ (\Leftrightarrow \hat v_i=-\delta^0_i)$, and we note that $\hat T^{ij}\,\hat k_j=0$. In addition when $T^{ij}$ in (\ref{29}) and (\ref{30}) is replaced by $\hat T^{ij}$ we can put $x=r\,\sin\theta\cos\phi$, $y=r\,\sin\theta\sin\phi$ and $z=r\,\cos\theta$, since we are taking the limit $r\rightarrow+\infty$, and this results in (\ref{29}) and (\ref{30}) being given by
\begin{align}
  \label{33}
  \frac{dP^i}{du}=(-\dot m, 0, 0, 0)\ ,
\end{align}
while $S^{ij}=0$ except for $S^{12}=-S^{21}$ (the $z$-component of the angular momentum of the matter distribution described by $T^{ij}$) which satisfies
\begin{align}
  \label{34}
  \frac{dS^{12}}{du} = \frac{d}{du}(m\,a)\ .
\end{align}
We see from (\ref{33}) that, since $P^i$ is the 4--momentum flowing \emph{away} from the system, the matter distribution described by $T^{ij}$ is losing energy provided $\dot m<0$, as one would expect. The angular momentum flowing away from the system is given by (\ref{34}).

\section{Discussion and conclusions}
\label{sec:disc}

The part (\ref{31}) of the energy-momentum-stress tensor which contributes to the evaluation of (\ref{29}) and (\ref{30}) giving the outward flow of 4--momentum and angular momentum at future null infinity consists of two parts
\begin{align}
  \label{35}
  t^{ij}_{(1)} = -\frac{2\,\dot m}{r^2}\,\hat k^i\,\hat k^j\ , \quad {\rm and}\quad
  t^{ij}_{(2)} = -\frac{3\,\dot m\,a}{r^3}(\hat k^i\,\hat\lambda^j+\hat k^j\,\hat\lambda^i)\ .
\end{align}
These tensors are defined in the neighbourhood of future null infinity which is the region of space-time with metric tensor $\eta_{ij}$. Therefore $t^{ij}_{(A)}=\eta^{ik}\,\eta^{jl}\,t_{(A)kl}$ for $A=1, 2$. It is interesting to note that they each satisfy the independent conservation equations 
\begin{align}
  \label{36}
  t^{ij}_{(1),j}=0\ ,\quad {\rm and} \quad
  t^{ij}_{(2),j}=0\ .
\end{align}
To verify this we note that the vectors $\hat k^i$ and $\hat\lambda^i$ are given by (\ref{32}) and in this region of space-time we have
\begin{align}
  \label{37}
  r^2=x^2+y^2+z^2\ .
\end{align}
Defining $\hat v^i=\delta^i_0$ we find the useful formulae
\begin{align}
  \label{38}
  \hat v_i\,\hat v^i=-1\ ,\quad
  \hat k_i\,\hat v^i=-1\ ,\quad
  \hat\lambda_i\,\hat v^i=0\ ,\quad
  \hat k_i\,\hat\lambda^i=0\ .
\end{align}
Therefore, we can show the following two results
\begin{align}
  \label{39}
  \hat k_{i,j} = \frac{1}{r}\,(\eta_{ij}+\hat k_i\,\hat v_j+\hat k_j\,\hat v_i-\hat k_i\,\hat k_j)
  \quad &\Rightarrow \quad
  \hat k_{i,j}\,\hat k^j=0
  \quad {\rm and}\quad
  \hat k^i{}_{,i}=\frac{2}{r}\ ,
  \\ \label{40}
  r_{,i}=\hat k_i-\hat v_i
  \quad &\Rightarrow\quad
  \ r_{,i}\,\hat k^i=1
  \quad {\rm and}\quad
  r_{,i}\,\hat\lambda^i=0\ .
\end{align}
These result are perhaps unsurprising when considering a large $r$ expansions in (\ref{22})--(\ref{24}). Finally, it is straightforward to show
\begin{align}
  \label{41}
  u_{,i}=-\hat k_i\ ,
\end{align}
which is a consequence of (\ref{25}) and
\begin{align}
  \label{42}
  \hat\lambda^i{}_{,j}\,\hat k^j=0\ ,\quad
  \hat\lambda^i{}_{,i}=0 \ .
\end{align}
Using these equations, it is straightforward to verify that (\ref{35}) satisfy (\ref{36}). Since $t^{ij}_{(A)}\,\hat k_j=0$ there is no flux of 4--momentum or angular momentum across the null cones $u={\rm constant}$ in this region of space-time and, since (\ref{36}) hold, this means that the 4--momentum or angular momentum that escapes to infinity is independent of $r$ for sufficiently large positive values of the radius $r$.

The key ingredient of our new approach to a Vaidya-type generalisation of the Kerr space-time was the introduction of the new variable $u=t-r$ where $r$ is the radial coordinate when working with Boyer-Lindquist coordinates. By promoting the mass and the angular momentum to become functions of $u$, we are able to construct the asymptotic energy-momentum-stress tensor. We find the asymptotic 4--momentum and angular momentum of the source and, as expected, our results reduce to the Vaidya generalisation of the Schwarzschild solution when considering $a \rightarrow 0$.

It follows from our Vaidya-type generalisation of the Kerr space-time that \emph{only asymptotically} (at future null infinity) is the resulting energy-momentum-stress tensor (a) is proportional to the square of a null vector $\hat k^i$ (the leading term in (\ref{31})) and (b) is the null vector $\hat k^i$ hypersurface orthogonal (which follows from (\ref{25})). Consequently we would expect that asymptotically the generalisation of the Kerr space--time described here would be useful as a ``background'' in a space-time perturbed by high frequency gravitational waves. This suggests an interesting topic for further study based on the approach to high frequency gravitational radiation in Kerr--Schild space-times described in an important paper by Taub \cite{Taub}.

\end{document}